# Predictive Analysis for Social Processes II: Predictability and Warning Analysis

Richard Colbaugh     Kristin Glass

*Abstract*—This two-part paper presents a new approach to predictive analysis for social processes. Part I identifies a class of social processes, called positive externality processes, which are both important and difficult to predict, and introduces a multi-scale, stochastic hybrid system modeling framework for these systems. In Part II of the paper we develop a systems theory-based, computationally tractable approach to predictive analysis for these systems. Among other capabilities, this analytic methodology enables assessment of process predictability, identification of measurables which have predictive power, discovery of reliable early indicators for events of interest, and robust, scalable prediction. The potential of the proposed approach is illustrated through case studies involving online markets, social movements, and protest behavior.

## I. INTRODUCTION

AS discussed in Part I of this two-part paper [1], predicting the outcome of social processes is both important and very challenging. Many social phenomena of interest in applications are positive externality processes (PEP), in which individuals are motivated to behave as others do. Research in the social and behavioral sciences provides compelling evidence that for such processes it is often not possible to obtain useful predictions using standard methods, which focus almost exclusively on the *intrinsic* characteristics of the process and its possible outcomes.

We propose that accurate prediction requires careful consideration of the interplay between the intrinsics of a process and the *social dynamics* which are its realization. We therefore adopt an inherently dynamical approach to predictive analysis: given a social process, a set of measurables, and the behavior of interest, we formulate prediction problems as questions about the reachability properties of the system.

As an illustrative example, consider the task of assessing the predictability of market share in a popular culture market, say for music, in which "buzz" about products spreads through various social networks. If, in a market containing two products with indistinguishable appeal, it is possible for one product to achieve a dominant market share, the market may be regarded to be unpredictable [2]. Conversely, in a predictable market the market shares of indistinguishable products evolve similarly and market shares of superior products are typically larger than those of inferior ones. In our formulation, market share dominance by product A is associated with a region of market share state space, and deciding whether A can achieve such dominance while possessing an appeal that is indistinguishable from product B is posed as a question about state space reachability.

More generally, in order to formulate prediction questions in terms of reachability, the behavior about which predictions are to be made is used to define system state space subsets of interest (SSI). Candidate measurables allow identification of indistinguishable starting sets (ISS), that is, sets of initial states and system parameters which cannot be resolved with the available data. This setup permits the four predictive analysis tasks of interest to us – predictability assessment, identification of useful measurables, warning analysis, and prediction – to be performed in a systematic manner. Predictability assessment involves determining which SSI can be reached from ISS and deciding if these reachability properties are compatible with the prediction goals. For example, if moving between state-parameter pairs within an ISS leads to unacceptably large variations in the probability of reaching the SSI, then the process is deemed unpredictable. This analysis leads naturally to a way of identifying those measurables with the most predictive power: these are the ISS coordinates for which predictability is most sensitive. If a system's reachability properties are incompatible with the prediction goals – if, say, "hit" and "flop" in a cultural market are both reachable from a single ISS – then the given prediction question should be refined in some way. Possible refinements include relaxing the level of detail to be predicted or introducing additional measurables.

If and when a predictable situation is obtained, the problems of discovering reliable early indicators for events of interest and forming robust predictions can be addressed. These problems are also readily studied within a reachability framework. Warning analysis involves identifying indicator (state) sets with the property that observing a trajectory entering an indicator set implies that the event of interest is likely to occur. Prediction entails estimating the probability that the process will evolve to an SSI and quantifying the uncertainty associated with this estimate.

The remainder of this paper transforms these intuitive notions into a rigorous, tractable methodology for predictive analysis, with a focus on predictability and early warning, and illustrates the utility of the approach through several real world case studies.

The research described in this paper was supported in part by the U.S. Department of Homeland Security and Sandia National Laboratories.

R. Colbaugh is with Sandia National Laboratories, Albuquerque, NM 87111 USA and New Mexico Institute of Mining and Technology, Socorro, NM 87801 USA (phone: 505-603-1248; e-mail: colbaugh@nmt.edu).

K. Glass is with New Mexico Institute of Mining and Technology, Socorro, NM 87801 USA (e-mail: kglass@icasa.nmt.edu).

## II. PREDICTIVE ANALYSIS

This section describes the proposed approach to predictive analysis for social processes. The presentation is structured to realize three objectives: 1.) provide reachability-based definitions for basic predictive analysis tasks; 2.) develop a rigorous, tractable methodology for reachability analysis; and 3.) derive efficient (reachability-based) algorithms for performing predictive analysis.

### A. Problem formulation

We now provide quantitative definitions for predictability assessment, identification of useful measurables, early warning, and robust prediction. Assume the behavior about which predictions are to be made and the measurables upon which these predictions can be based have been used to specify the system SSI and ISS, respectively. Denote by $\Sigma_{sp}$ the social process of interest, and suppose it is modeled using the stochastic hybrid system (S-HDS) framework developed in [1].

**Definition 2.1:** Let $X \subseteq \Re^n$ and $Par \subseteq \Re^p$ denote the (bounded) state and parameter sets for social process $\Sigma_{sp}$, $P_0$ be a subset of Par, and $X_0$, $X_{s1}$, $X_{s2}$ be subsets of X. Suppose $X_0 \times P_0$ and $\{X_{s1}, X_{s2}\}$ are the ISS and SSI, respectively, corresponding to the prediction question. Let a specification $\delta$ be given for the acceptable level of variation or uncertainty in system behavior relative to $\{X_{s1}, X_{s2}\}$, and suppose $(x_0^*, p_0^*) \in X_0 \times P_0$ is the best estimate for the process initialization. *Initial state (IS) predictability assessment* involves determining whether unacceptable variation in the reachability properties of $\{X_{s1}\}$ results from changing the system initialization. More precisely, a prediction problem is *IS unpredictable* if $\gamma_{max} - \gamma_{BE} > \delta$, where $\gamma_{max}$ and $\gamma_{BE}$ are the probabilities of $\Sigma_{sp}$ reaching $X_{s1}$ from $X_0 \times P_0$ and $(x_0^*, p_0^*)$, respectively; the problem is IS predictable otherwise. *Eventual state (ES) predictability assessment* involves evaluating the acceptability of uncertainty in system behavior relative to reachability of the two sets $X_{s1}$ and $X_{s2}$. Thus a situation is *ES unpredictable* if $\min(\gamma_1, \gamma_2) > \delta$, where $\gamma_i$ is the probability of $\Sigma_{sp}$ reaching $X_{si}$, and is ES predictable otherwise.

Note that in ES predictability problems it is expected that the two sets $\{X_{s1}, X_{s2}\}$ represent qualitatively different system behaviors (e.g., hit and flop in a cultural market), so that if the probability of reaching each from $X_0 \times P_0$ is relatively high then system behavior is unpredictable in an application-relevant sense.

These predictability concepts form the basis for our definition of useful measurables:

**Definition 2.2:** Let the components of the vectors $(x_0, p_0) \in X_0 \times P_0$ which comprise the ISS be denoted $x_0 = [x_{01} \ldots x_{0n}]^T$ and $p_0 = [p_{01} \ldots p_{0p}]^T$. The *measurables with most predictive power* correspond to the state variables $x_{0j}$ and/or parameters $p_{0k}$ for which predictability is most sensitive.

We do not specify a particular measure of sensitivity to be used in identifying measurables with maximum predictive power and do not require that these measurables actually have sufficient power to be useful. Such considerations are ordinarily application-dependent and are explored in [3].

**Remark 2.1:** Definitions 2.1 and 2.2 characterize the role played by *initial* states in the predictability of social processes. In some cases it is useful to expand this formulation to allow consideration of states other than initial states. For instance, we will show in the case studies that very early time series are often predictive for PEP, suggesting that it can be valuable to consider initial state *trajectory segments*, rather than just initial states, when assessing predictability. This extension can be accomplished by redefining the ISS $X_0 \times P_0$, for example by augmenting the state space X with an explicit time coordinate.

We now consider warning analysis and prediction.

**Definition 2.3:** Let an event of interest be specified in terms of some SSI $X_s$ (e.g., $\Sigma_{sp}$ reaching or leaving $X_s$) and let the required warning accuracy be given (e.g., a warning signal is to be issued only if the probability of event occurrence exceeds some level). *Warning analysis* involves identifying one or more state-parameter subsets $X_w \times P_w \subseteq X \times Par$, which we term *indicators*, with the property that "observing an indicator" – that is, observing the system trajectory entering $X_w \times P_w$ – corresponds to the issuing of a warning with the specified accuracy.

For instance, we often specify the warning accuracy and indicator in such a way that if the indicator is observed then the probability of event occurrence exceeds the given threshold. Note that this definition for warning analysis and warning indicators captures the essence of the informal usage of these terms and is also convenient for formal analysis.

Finally, we have:

**Definition 2.4:** Let a behavior of interest be specified in terms of the trajectories of the social process $\Sigma_{sp}$ (e.g., the ultimate state for a convergent process or the maximum value attained by a state variable) and let $X_0 \times P_0$ denote the ISS. *Prediction* entails 1.) estimating the salient characteristics of the behavior and 2.) quantifying the uncertainty associated with this estimate.

This definition specifies that we seek "best estimate plus uncertainty" predictions, for example estimating the state $x^*$ to which $\Sigma_{sp}$ will converge given our best guess for $(x_0^*, p_0^*) \in X_0 \times P_0$, and also computing the set of all possible values for $x^*$ associated with feasible pairs $(x_0, p_0) \in X_0 \times P_0$.

### B. Stochastic reachability analysis

The previous section formulates predictive analysis problems as reachability questions. In this section we show that these reachability questions can be addressed by adopting an analysis methodology which is related to familiar Lyapunov function stability analysis [4,5]. More specifically, we seek a scalar function of the system state that permits conclusions to be made regarding reachability *without computing system trajectories*. We refer to these as "altitude functions" to provide an intuitive sense of their role in reachability analysis: if some measure of "altitude" is low on the ISS and high on an SSI, and if the expected rate of change of altitude along system trajectories is nonincreasing, then it is unlikely for trajectories to reach this SSI from the ISS.

Part I of this paper [1] develops an S-HDS framework for modeling a broad range of social processes, including PEP, and we employ that framework here. Let $\Sigma_{\text{S-HDS}}$ denote a general S-HDS with bounded state space $Q \times X$, and suppose that the dynamics of $\Sigma_{\text{S-HDS}}$ is characterized by the infinitesimal generator $BA(q,x)$ [6]. We quantify the uncertainty associated with $\Sigma_{\text{S-HDS}}$ by specifying *bounds* on the possible values for some system parameters and perturbations and *probabilistic descriptions* for other uncertain system elements and disturbances. Given this representation for social processes, it is natural to seek a probabilistic assessment of system reachability.

We begin with an investigation of (probabilistic) reachability on infinite time horizons. The following result is proved in [7] and is instrumental in our development:

**Lemma 2.1:** Consider a stochastic process $\Sigma_s$ with bounded state space X, and let $\underline{x}(t)$ denote the stopped process associated with $\Sigma_s$ (i.e., $\underline{x}(t)$ is the trajectory of $\Sigma_s$ which starts at $x_0$ and is stopped if it encounters the boundary of X). If $A(\underline{x}(t))$ is a nonnegative supermartingale then for any $x_0$ and $\lambda > 0$

$$P\{\sup A(\underline{x}(t)) \geq \lambda \mid \underline{x}(0) = x_0\} \leq A(x_0) / \lambda.$$

Denote by $X_0 \subseteq X$ and $X_u \subseteq X$ the initial state set and SSI, respectively, for the continuous system component of $\Sigma_{\text{S-HDS}}$, and assume that X and the parameter set $Par \subseteq \mathfrak{R}^p$ are both bounded. Thus, for instance, the SSI is a subset of the continuous system state space X alone; this is typically the case in applications and is easily extended if necessary. We are now in a position to state our first result:

**Theorem 1:** $\gamma$ is an upper bound on the probability of trajectories of $\Sigma_{\text{S-HDS}}$ reaching $X_u$ from $X_0$ while remaining in $Q \times X$ if there exists a family of differentiable functions $\{A_q(x)\}_{q \in Q}$ such that

- $A_q(x) \leq \gamma$ $\forall x \in X_0$, $\forall q \in Q$;
- $A_q(x) \geq 1$ $\forall x \in X_u$, $\forall q \in Q$;
- $A_q(x) \geq 0$ $\forall x \in X$, $\forall q \in Q$;
- $BA_q(x) \leq 0$ $\forall x \in X$, $\forall q \in Q$, $\forall p \in Par$.

**Proof:** As $BA_q(x)$ is the infinitesimal generator for $\Sigma_{\text{S-HDS}}$, the third and fourth conditions of the theorem imply that $A(q(t),x(t))$ is a nonnegative supermartingale $\forall p \in Par$. Thus, from Lemma 2.1, we can conclude that $P\{x(t) \in X_u$ for some $t\} \leq P\{\sup A(q(t),x(t)) \geq 1 \mid x(0)=x_0\} \leq A(q,x_0) \leq \gamma$ $\forall x_0 \in X_0$, $\forall q \in Q$, $\forall p \in Par$. ∎

**Remark 2.2:** Theorem 1 extends a similar result given in [4] to allow probability bounds to be established for reachability questions in the presence of set-bounded uncertainties.

The preceding result characterizes reachability of S-HDS on infinite time horizons. In some situations, including important applications involving social systems, it is of interest to study system behavior on *finite* time horizons. The following result is useful for such analysis:

**Theorem 2:** $\gamma$ is an upper bound on the probability of trajectories of $\Sigma_{\text{S-HDS}}$ reaching $X_u$ from $X_0$ during time interval [0,T], while remaining in $Q \times X$, if there exists a family of differentiable functions $\{A_q(x,t)\}_{q \in Q}$ such that

- $A_q(x,t) \leq \gamma$ $\forall (x,t) \in X_0 \times 0$, $\forall q \in Q$;
- $A_q(x,t) \geq 1$ $\forall (x,t) \in X_u \times [0,T]$, $\forall q \in Q$;
- $A_q(x,t) \geq 0$ $\forall (x,t) \in X \times \mathfrak{R}^+$, $\forall q \in Q$;
- $BA_q(x,t) \leq 0$ $\forall (x,t) \in X \times \mathfrak{R}^+$, $\forall q \in Q$, $\forall p \in Par$.

**Proof:** The proof follows immediately from that of Theorem 1 once it is observed that $P\{\underline{x}(t) \in X_u$ for some $t \in [0,T]\} = P\{(\underline{x}(t),t) \in X_u \times [0,T]\}$. ∎

The idea for the proof of Theorem 2 was suggested in [8].

The analytic methodology employed above also can be used to determine *lower bounds* on the probability of reaching an SSI. Consider a stochastic process $\Sigma_s$ with bounded state space X and SSI $X_u \subseteq X$, and suppose it is of interest to determine a lower bound on the probability of $\Sigma_s$ reaching $X_u$ during some time interval [0,T]. Assume, for simplicity, that the dynamics of $\Sigma_s$ is such that $X_u$ is invariant (i.e., if $\Sigma_s$ enters $X_u$ it cannot escape this set); this situation is common in applications.

We formulate the problem in terms of *escaping* $X_e = X \setminus X_u$, as the probability of reaching $X_u$ is identical to that of escaping $X_e$, and suppose that $x_0 \in X_e$ (otherwise $x_0 \in X_u$ and the problem is trivial). Let $X^* = X_e \times T$ denote the set obtained by augmenting $X_e$ with the time value $t = T$. Observe that

$P\{(\underline{x}(t),t) \in X^*\} = P\{\underline{x}(t) \in X_e$ at $t=T\} = P\{\underline{x}(t) \in X_e$ $\forall t \in [0,T]\}$,

because once a trajectory escapes $X_e$ it cannot return to this set. Now an upper bound $\gamma$ can be determined for the probability of reaching $X^*$, $P\{(\underline{x}(t),t) \in X^*\} \leq \gamma$, using the results developed in the preceding discussion. Then, since

$P\{\underline{x}(t) \in X_e$ $\forall t \in [0,T]\} + P\{\underline{x}(t^*) \notin X_e$ for some $t^* \in [0,T]\} = 1$

we can conclude

$P\{\underline{x}(t^*) \in X_u$ for $t^* \in [0,T]\} = P\{\underline{x}(t^*) \notin X_e$ for $t^* \in [0,T]\} \geq 1 - \gamma$.

Thus we have proved

**Theorem 3:** Suppose $\gamma$ is an upper bound on the probability of $\Sigma_s$ reaching $X^* = X_e \times T$. Then $\gamma_{lb} = 1 - \gamma$ is a lower bound on the probability of $\Sigma_s$ reaching $X_u$ during the time interval [0,T].

The preceding theoretical results are of direct practical interest only if it is possible to efficiently compute families of altitude functions $\{A_q(x)\}_{q \in Q}$. Toward that end, observe that the results presented in Theorems 1-3 specify *convex* conditions to be satisfied by the associated altitude functions. Thus the search for altitude functions can be formulated as a convex programming problem [9]. Moreover, if the system of interest admits a polynomial description (i.e., the system vector fields are polynomials and system sets are semialgebraic) and if we restrict our search to polynomial altitude functions, then the search can be carried out using sum of squares (SOS) optimization [4,10]. Importantly, this approach is tractable: for fixed polynomial degrees, the computational complexity of the associated SOS program grows polynomially in the dimension of the continuous state space, the cardinality of the discrete state set, and the dimension of the parameter space

### C. Reachability-based predictive analysis

Having formulated predictive analysis for social processes in terms of system reachability and presented a methodology for assessing reachability, we are now in a position to derive algorithms for predictive analysis. In what follows we focus on the tasks of predictability assessment and early warning analysis; algorithms for identifying measurables with predictive power and forming predictions are developed in [11].

Consider first predictability assessment as characterized in Definition 2.1. We have the following algorithms for infinite time horizon predictability:

**Algorithm 2.1: IS predictability** (outline)
Given: social process of interest is $\Sigma_{\text{S-HDS}}$, ISS = $X_0 \times P_0$, best estimate for initialization is $(x_0^*, p_0^*) \in X_0 \times P_0$, SSI = $X_s$, and acceptable level of variation = $\delta$.
Procedure:
- compute (upper bound for) probability $\gamma_{\max}$ of $\Sigma_{\text{S-HDS}}$ reaching $X_s$ from $X_0 \times P_0$;
- compute (upper bound for) probability $\gamma_{\text{BE}}$ of $\Sigma_{\text{S-HDS}}$ reaching $X_s$ from $(x_0^*, p_0^*)$;
- if $\gamma_{\max} - \gamma_{\text{BE}} > \delta$ then problem is IS unpredictable, else problem is IS predictable.

Note: $\gamma_{\max}$, $\gamma_{\text{BE}}$ can be computed using Theorem 1 and SOS programming.

**Algorithm 2.2: ES predictability** (outline)
Given: social process of interest is $\Sigma_{\text{S-HDS}}$, ISS = $X_0 \times P_0$, SSI = $\{X_{s1}, X_{s2}\}$, and acceptable level of uncertainty = $\delta$.
Procedure:
- compute (upper bound for) probability $\gamma_1$ of $\Sigma_{\text{S-HDS}}$ reaching $X_{s1}$ from $X_0 \times P_0$;
- compute (upper bound for) probability $\gamma_2$ of $\Sigma_{\text{S-HDS}}$ reaching $X_{s2}$ from $X_0 \times P_0$;
- if $\min(\gamma_1, \gamma_2) > \delta$ then problem is ES unpredictable, else problem is ES predictable.

Note: $\gamma_1$, $\gamma_2$ can be computed using Theorem 1 and SOS programming.

**Remark 2.3:** IS and ES predictability can be assessed on finite time horizons in the same manner, with the required probability bounds being computed using SOS programming and the criteria given in Theorem 2.

We now examine the warning analysis problem specified in Definition 2.3. Assume given a social process $\Sigma_{\text{sp}}$, some SSI $X_s \subseteq X$, a finite time interval $[0,T]$, and a warning accuracy $\alpha \in (0,1]$. We are interested in two versions of the problem, corresponding to whether the event of interest involves $\Sigma_{\text{sp}}$ *reaching* $X_s$ or *escaping from* $X_s$. In either case, the warning is to be issued if and only if the probability of event occurrence during the time interval $[0,T]$ following this warning is at least $\alpha$. Following Definition 2.3, we seek an "indicator" $X_w \subseteq X$ with the property that if $\Sigma_{\text{sp}}$ enters $X_w$ then the probability of event occurrence is at least $\alpha$.

Consider first the situation in which the event to be anticipated is $\Sigma_{\text{sp}}$ reaching $X_s$. In this case, the objective is to identify the largest $X_w$, with $X_s \subseteq X_w$ necessarily, such that $x(0) \in X_w$ implies $P\{x(t) \in X_s \text{ for some } t \in [0,T]\} \geq \alpha$. Suppose, as above, that the dynamics of $\Sigma_{\text{sp}}$ is such that $X_s$ is invariant. The following algorithm provides a solution to this version of the warning problem:

**Algorithm 2.3: reach warning analysis** (outline)
Given: social process of interest is $\Sigma_{\text{sp}}$, SSI = $X_s$, and warning accuracy = $\alpha$.
Procedure:
Initialize $X_{w0} = X_s$.
For k = 1, 2, …, K:
- Incrementally enlarge $X^*_{wk} \supseteq X_{w(k-1)}$.
- Compute $\gamma_{lb} \leq P\{x(t) \in X_s \text{ for some } t \in [0,T] \mid x(0) \in X^*_{wk}\}$ (via Theorem 3 and SOS programming).
- If $\gamma_{lb} \geq \alpha$ set $X_{wk} = X^*_{wk}$ and RETURN, else STOP.

Note: There exist numerous methods for incrementally "growing" a sequence of nesting sets $X_{wk}$ [e.g., 12].

Next consider the case in which the event of interest is $\Sigma_{\text{sp}}$ escaping from $X_s$. Here the goal is to identify the largest $X_w$, with $X \setminus X_w \subseteq X_s$ necessarily, such that $x(0) \in X_w$ implies $P\{x(t) \notin X_s \text{ for some } t \in [0,T]\} \geq \alpha$. Suppose that $X \setminus X_s$ is invariant for the dynamics of $\Sigma_{\text{sp}}$. The next algorithm provides a solution to the escape warning problem:

**Algorithm 2.4: escape warning analysis** (outline)
Given: social process of interest is $\Sigma_{\text{sp}}$, SSI = $X_s$, and warning accuracy = $\alpha$.
Procedure:
Initialize $X_{w0} = X \setminus X_s$.
For k = 1, 2, …, K:
- Incrementally enlarge $X^*_{wk} \supseteq X_{w(k-1)}$.
- Compute $\gamma_{lb} \leq P\{x(t) \in X \setminus X_s \text{ for some } t \in [0,T] \mid x(0) \in X^*_{wk}\}$ (via Theorem 3 and SOS programming).
- If $\gamma_{lb} \geq \alpha$ set $X_{wk} = X^*_{wk}$ and RETURN, else STOP.

An alternative approach to identifying a warning indicator set $X_w$ for a given social process-event pair is to compare trajectories of $\Sigma_{\text{sp}}$ which lead to the event of interest with those that do not. If differences are found between the dynamics of the two classes of processes, and if these differences can be expressed in terms of some $X_w$, then this analysis identifies an empirically-grounded event indicator. Stan-dard statistical or machine learning classification methods can be employed for this task in cases where there are sufficient data to enable an empirical comparison. For social processes, however, it is often the case that available data consists mainly of "positive" events, involving trajectories which lead to the event of interest. In such situations, it is sometimes possible to construct useful *synthetic* ensembles of negative events and to formulate a comparison study by combining the actual positive instances and the synthetic negative instances. This approach to warning analysis can be effective if good models are available for building the synthetic ensembles.

### III. CASE STUDIES

This section presents three case studies involving predictive analysis for classes of social processes that have proven to be both practically important and challenging to predict. We begin with a discussion of predictability assessment for online markets and then address early warning analysis for social movements and mobilization/protest events.

#### A. Online markets

Consider an online market in which individuals visit a web site, browse an assortment of available items, and choose one or more items to download. An interesting and surprising characteristic of these markets – and many other markets as well – is that they are often both unequal and unpredictable: a few items capture a large share of the market, but which items achieve popularity appears to be hard to anticipate. For instance, the study reported in [2] created an artificial music market and demonstrated this phenomenon experimentally. Moreover, that work showed that increasing the opportunity for social influence increased both the inequality of the ultimate market shares and the unpredictability of which songs attained market dominance. Our study of CNET, the online software library, yielded similar results [3]. The positive externalities present in these markets makes predictive analysis using standard methods a challenging undertaking.

We now assess the feasibility of forecasting ultimate market share in online markets. Consider a market visited by a sequence of consumers, with each visitor choosing between two items {A, B}; generalizing this simple binary choice setting to any finite number of choices is straightforward. We model this situation by supposing that agent i chooses item A with probability

$$\Sigma_{online} \qquad P_i(A) = \beta\pi + (1-\beta) f$$

where $f \in [0,1]$ is item A's current market share, $(1-\beta)$ quantifies the intensity of social influence (with $\beta \in [0,1]$), and $\pi$ is the probability of an agent choosing A in the "no social influence" case (i.e., when $\beta=1$). Agent i selects item B with probability $1 - P_i(A)$. In this model, $\pi$ can be interpreted as a measure of the "appeal" of item A (relative to B), f is the social signal, and $\beta$ quantifies the relative importance of appeal and social influence in the decision-making process.

The model $\Sigma_{online}$ is extremely simple, perhaps the simplest possible representation which captures the effects of both social influence and appeal in an online market. Nevertheless, this model is able to reproduce the key behaviors observed in the music market study described in [2], in our investigation of CNET site dynamics, and in other online markets (e.g., for books and DVDs) [3]. In particular, as social influence (SI) increases ($\beta$ decreases) both inequality and unpredictability of market shares increase. Thus, despite its simplicity, $\Sigma_{online}$ provides a useful starting point for studying predictability of online markets. Note that $\Sigma_{online}$ can be written in the form of the continuous system portion of the S-HDS model $\Sigma_{S-HDS}$, with state variables $x_1 = f$ and $x_2 = 1/(t+1)$. Consequently, the system's reachability properties can be determined using Theorems 1-3.

We now investigate the predictability of ultimate market share for the system $\Sigma_{online}$. The standard approach to market share prediction is to assume that item appeal is a relevant measurable, estimate appeal in some way, and use this estimate to predict market share. To examine the utility of this approach, we assess ES predictability of market share for items with identical appeal ($\pi=1/2$) and identical initial market shares (f(0)=1/2). If it is reasonably likely that the market will evolve so one or the other item dominates (f becomes large or small), then the market dynamics is not very dependant on item appeal and therefore is unpredictable using the standard approach. In this case we should seek a different prediction method, perhaps based on other measurables. Alternatively, if market dominance by either item is unlikely then the market dynamics depends on item appeal in a more predictable way and the standard method may be useful.

We evaluate ES predictability, as specified in Definition 2.1, via Algorithm 2.2 and Theorem 1. Let the two SSI, $X_{s1}$ and $X_{s2}$, be defined to correspond to, respectively, $f \approx 1/2$ (approximately equal market share) and large/small f (market dominance by one or the other item). Define the ISS $X_0$ to be a small set surrounding $f(0) = 1/2$, the identical initial market share condition. Then, if both $X_{s1}$ and $X_{s2}$ are likely to be reached from $X_0$, the problem is ES unpredictable (and also unpredictable from a practical viewpoint). See Figure 1 for a diagram depicting the basic setup.

As an illustration of the insights obtainable with such analysis, consider the high SI case corresponding to small $\beta$ in $\Sigma_{online}$. For a broad range of noise models, the analysis generates relatively high probability bounds for reachability of both $X_{s1}$ and $X_{s2}$ from $X_0$ (e.g., $\gamma \in [0.3, 0.4]$ is typical). Thus two qualitatively different outcomes – market share equity ($X_{s1}$) and market shares dominance ($X_{s2}$) – are both likely, indicating that the system is ES unpredictable. This result is consistent with empirical findings [e.g., 2] and suggests that the standard approach to market share prediction is not likely to produce accurate forecasts.

Next consider the problem of searching for alternative measurables which provide better predictability properties in the high SI case. For example, it might be supposed that very early market share time series data would be useful for prediction when SI is high. The intuition behind this idea is that the "herding" behavior that can arise from SI, and which makes market prediction hard using standard methods, may lead to a lock-in effect, in which very early market share leaders become difficult to displace. To test this hypothesis, define $X_0^*$ to be a small set surrounding $f(t^*) = 1/2$, where $t^*$ is a small *but nonzero* time (see Figure 1). We compute, using Theorem 1 and SOS programming, an upper bound on the probability that $\Sigma_{online}$ with $\pi=1/2$ will evolve from $X_0^*$ to $X_{s1}$ and $X_{s2}$. In this case, the analysis generates large upper bounds for the probability of reaching $X_{s1}$ and small bounds for the probability of reaching of $X_{s2}$ (typical bounds are on the order $\gamma\sim 0.9$ and $\gamma\sim 10^{-3}$, respectively). Thus using very early time series data to refine the ISS produces a more predictable situation, in which indistinguishable market configurations evolve to indistinguishable outcomes.

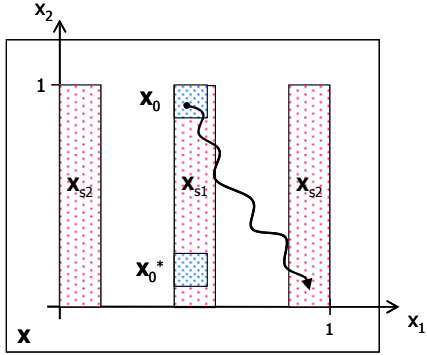

Fig. 1. Setup for online market predictability assessment.

### B. Social movements

Social movements are large, informal groupings of individuals and/or organizations focused on a particular issue, for instance of political, social, economic, or religious significance. There is considerable interest to develop methods for distinguishing successful social movements, that is, movements which attract significant followings, from unsuccessful ones early in their lifecycle. This task is naturally cast as a warning problem within the proposed approach to predictive analysis. We study the problem in two phases: 1.) a *theoretical investigation*, in which a collection of general models for social movement dynamics are analyzed, and 2.) an *empirical study*, involving the emergence and diffusion of Sweden's Social Democratic Party (SDP).

We begin with the theoretical investigation. Movement success is quantified by defining an SSI, $X_s$, that corresponds to a level of movement membership consistent with movement goals, and we seek to identify an indicator set $X_w$ which permits early recognition of those movements that are likely to evolve to $X_s$ (see Definition 2.3). We construct a class of social movement models within a diffusion of innovations framework [11]; the resulting models are of the general form $\Sigma_{S-HDS}$ developed in Part I of this paper [1] and are consistent with the social movement theory (SMT) literature [e.g., 13]. In particular, the models capture important structural features of social networks, including the existence and topology of *social contexts*, that is, localized social settings defined by work, family, or physical neighborhood within which close interactions take place [1].

We conduct reach warning analysis by employing the procedure outlined in Algorithm 2.3. Briefly, the theoretical study produced two main results. First, the degree to which movement-related activity shows *early diffusion* across multiple social contexts is a powerful distinguisher of successful and unsuccessful social movements. Indeed, this measurable has considerably more predictive power than the *magnitude* of such activity and also more power than various system intrinsics. Second, large social movements occur with finite probability only if 1.) the intra-context "infectivity" of the movement exceeds a certain threshold, and 2.) the inter-context interactions associated with the movement take place with a frequency that is larger than another threshold.

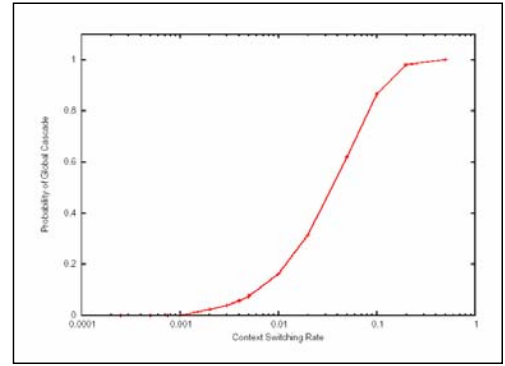

Fig. 2. Probability of global diffusion of social movement as a function of context switching rate.

The latter result is particularly interesting, as it is reminiscent of, and significantly extends, well-known results for epidemic thresholds in disease propagation models. For instance, the characterization of intra-context infectivity generalizes the notion of epidemic *reproduction number* [11] to social movements. More intriguing is the completely new condition on inter-context interactions: in order for a social movement to propagate "globally", that is, to extend into social contexts beyond its original local setting, the probability of context interaction must exceed a threshold value. This threshold behavior is depicted in Figure 2, which shows the way the probability of realizing global propagation depends on the rate at which individuals interact across social contexts; it can be seen that this dependency exhibits a classic threshold behavior. Note that the probabilities shown in Figure 2 are provably-correct upper bounds for the global cas-cade probabilities and were obtained using Theorem 1 and SOS programming.

The empirical investigation of early warning analysis for social movements focuses on the emergence and growth of the Swedish SDP. The case of the SDP is particularly relevant for our purposes, as the early activities of political "agitators" associated with the SDP led to the establishment of a well-defined and well-documented network linking previously disparate geographically- and demographically-based social contexts in Sweden [13]. We explore the role played by this inter-context network by analyzing archived data [14] and published accounts describing the dynamics of the SDP. Our investigation uses standard time series analysis techniques similar to those employed in [13], and reveals that an important predictor of SDP spatio-temporal dynamics is *early diffusion* of SDP-related activity across geographically-based social contexts. Thus both the theoretical and empirical investigations suggest that early social network dynamics are critical to social movement success.

### C. Mobilization / protest

This case study examines whether diffusion across social contexts is a useful early indicator for successful mobilization and protest events. The investigation focuses on Muslim reaction to six recent incidents, each of which appeared at the outset to have the potential to trigger significant protests:

- publication of photographs and accounts of prisoner abuse at Abu Ghraib in Spring 2004;
- publication of cartoons depicting Mohammad in the Danish newspaper *Jyllands-Posten* in September 2005;
- distribution of the DVD "I was blind but now I can see" in Egypt in October 2005;
- the lecture given by Pope Benedict XVI in September 2006 quoting controversial material concerning Islam;
- Salman Rushdie being knighted in June 2007;
- republication of the "Danish cartoons" in various newspapers in February 2008.

Recall that the first Danish cartoons event ultimately led to substantial Muslim mobilization, including massive protests and considerable violence, and that the Egypt DVD event also resulted in significant Muslim protest and violence. In contrast, Muslim outrage triggered by Abu Ghraib, the pope lecture, the Rushdie knighting, and the second Danish cartoons event all subsided quickly with essentially no violence. Therefore, taken together, these six events provide a useful setting for testing whether the extent of early diffusion across social contexts can be used to distinguish nascent mobilization events which become large and self-sustaining (and potentially violent) from those that quickly dissipate.

A central element in the proposed approach to early warning analysis is the measurement, and appropriate processing, of social dynamics associated with the process of interest. In the present case study we use *online* social activity as a proxy for real world diffusion of mobilization-relevant information. More specifically, we use blog communications and discussions as our primary data set. The "blogosphere" is modeled as a graph composed of two types of vertices, the blogs themselves and the concepts which appear in them. Two blogs are linked if a post in one hyperlinks to a post in the other, and a blog is linked to a concept if the blog contains (significant) occurrences of that concept. Among other things, this blog graph model enables the identification of blog communities – that is, groups of blogs with intra-group edge densities that are significantly higher than expected [15]. In what follows, these blog communities serve as one proxy for social contexts.

Consider the problem of deriving early indicators that reliably distinguish successful and unsuccessful mobilization/ protest events. We adopt an approach which is analogous to that used in the preceding case study, quantifying mobilization success in terms of an SSI $X_s$ that is "large enough", and seeking to identify an indicator condition that permits early recognition of events likely to evolve to $X_s$. In this case study, however, we employ the second of the two methods for identifying warning indicators given in Section II.C. Thus the warning condition is derived by comparing trajectories of the social process which led to successful mobilization events with those that did not. Because in the present application the available data is insufficient to support a purely empirical analysis, these data are augmented through construction of *synthetic* ensembles of events. This approach is feasible because the social diffusion model $\Sigma_{S\text{-HDS}}$ presented in [1] provides a reasonable mechanism for generating these ensembles.

More specifically, the following procedure is proposed for mobilization/protest warning analysis using blog data:

Given a potential triggering event of interest

1. Use key words and concepts associated with the triggering event to collect relevant blog posts and build the associated blog graph.
2. Identify the relevant blog social contexts (e.g., graph community-based, language-based).
3. Assemble post volume time series for each social context and compute post/context entropy (PCE) time series associated with the post volume time series.
4. Construct a synthetic ensemble of PCE time series from (actual) post volume dynamics using the S-HDS social diffusion model $\Sigma_{S\text{-HDS}}$ [1].
5. Perform motif detection: compare the actual PCE time series to the synthetic ensemble series to determine if the early diffusion of activity across contexts is "excessive".

We now provide a few additional details concerning this procedure. Step 1 is by now standard, and various off-the-shelf tools exist which can perform this task. In Step 2 we use two definitions for blog social context: graph-based, in which contexts are graph communities identified through community extraction applied to the blog graph [e.g., 15], and language-based, in which contexts are defined based on the language of the posts. In Step 3, post volume for a given context i and sampling interval t is obtained by counting the number of relevant posts made in the blogs comprising context i during interval t. PCE for a given sampling interval t is defined as follows: $PCE(t) = -\Sigma_i f_i(t) \log(f_i(t))$, where $f_i(t)$ is the fraction of total relevant posts made during interval t which occur in context i.

Given the post volume time series obtained in Step 3, Step 4 involves the construction of an ensemble of PCE time series which would be expected under "normal circumstances", that is, if Muslim reaction to the triggering event diffused from a small "seed set" of initiators according to SMT social dynamics. For this study, we use the multi-scale social diffusion model $\Sigma_{S-HDS}$ given in [1] to generate the PCE time series ensembles. Finally, motif detection in Step 5 is carried out by searching for time periods, if any, during which the actual PCE time series exceeds the mean of the synthetic PCE ensemble by at least two standard deviations.

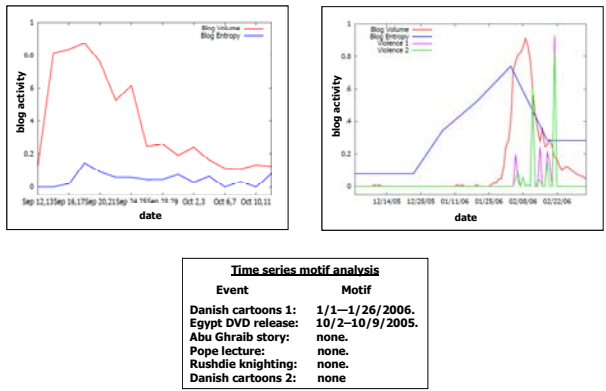

Fig. 3. Sample results for Islamic mobilization case study. The time series plots at the top correspond to the pope event (left) and first Danish cartoons event (right). In each plot, the red curve is blog volume and the blue curve is blog entropy; the Danish cartoon plot also shows two measures of violence (cyan and magenta curves). Note that while the data are scaled to allow multiple data sets to be graphed on each plot, the scale for entropy is consistent across plots to enable cross-event comparison. The table at the bottom summarizes the results of the motif analysis study.

Sample results of applying the proposed approach to early warning analysis to the Islamic mobilization case study are shown in Figure 3. It can be seen that early diffusion of discussions across blog communities is, indeed, an indicator that the associated Islamic mobilization event will be large. Such diffusion is observed in the mobilization associated with the first Danish cartoons and Egypt DVD events and not with the other four events, and this early diffusion is excessive relative to the synthetic ensemble. More specifically, in the case of the first Danish cartoons event, the PCE of relevant discussions (blue curve) experiences a dramatic increase a few weeks before the corresponding increase in volume of blog discussions (red curve); this latter increase, in turn, takes place before any violence (see Figure 3). In contrast, in the case of the pope event, PCE of blog discussions is small relative to the cartoons event, and any increase in this measure lags discussion volume. Similar curves are obtained for the other four events. More importantly, the proposed motif detection process also yields the expected result: motifs are found only for the Danish cartoons and Egypt DVD events, and these motifs precede significant blog volume and real world violence. Note that qualitatively similar results are obtained for the graph community-based and language-based definitions of social context. This case study suggests that early diffusion of mobilization-related activity (here blog discussions) across disparate social contexts may be a useful early indicator of successful mobilization events.